\documentclass[letterpaper, 10 pt, conference]{ieeeconf}  

\IEEEoverridecommandlockouts                              
\usepackage{graphicx} 
\usepackage{amsmath}
\usepackage{array}
\usepackage{graphicx}
\usepackage{float}
\usepackage{caption}
\usepackage{subcaption}
\usepackage{booktabs} 
\usepackage{tabularx}
\usepackage{bm} 
\usepackage{relsize}
\usepackage{adjustbox} 
\usepackage{textcomp}
\usepackage{multicol}
\usepackage{multirow}
\usepackage{url}

\usepackage{times} 
\usepackage{amsmath} 
\usepackage{amssymb}  

\usepackage{authblk}

\title{\LARGE \bf
Redditors in Recovery: Text Mining Reddit to Investigate Transitions into Drug Addiction
}
\author[1]{John Lu} %
\author[2]{Sumati Sridhar} %
\author[3]{Ritika Pandey} %
\author[4]{Mohammad Al Hasan} %
\author[5]{George Mohler} %

\affil[1]{Department of Computer Science. University of California, Irvine} %
\affil[2]{Department of Statistics and Operations Research. University of North Carolina, Chapel Hill}
\affil[3,4,5]{Department of Computer and Information Science, IUPUI}
\makeatletter
\renewcommand\AB@affilsepx{, \protect\Affilfont}
\makeatother
\affil[1]{johnl19@uci.edu}
\affil[2]{sumati@live.unc.edu} 
\affil[3]{ripande@iu.edu}
\affil[4]{alhasan@iupui.edu}
\affil[5]{gmohler@iupui.edu}

\begin{document}
\maketitle
\thispagestyle{plain}
\pagestyle{plain}

\begin{abstract}
Increasing rates of opioid drug abuse and heightened prevalence of online support communities underscore the necessity of employing data mining techniques to better understand drug addiction using these rapidly developing online resources. In this work, we obtain data from Reddit, an online collection of forums, to gather insight into drug use/misuse using text data from users themselves. Specifically, using user posts, we trained 1) a binary classifier which predicts transitions from casual drug discussion forums to drug recovery forums and 2) a Cox regression model that outputs likelihoods of such transitions. In doing so, we found that utterances of select drugs and certain linguistic features contained in one's posts can help predict these transitions.  
Using unfiltered drug-related posts, our research delineates drugs that are associated with higher rates of transitions from recreational drug discussion to support/recovery discussion, offers insight into modern drug culture, and provides tools with potential applications in combating the opioid crisis. 
\end{abstract}

\section*{INTRODUCTION}
The rate of nonmedical opioid use has increased markedly since the early 2000s. While recent efforts have been made to curb over-prescribing \cite{painManagement,OhioCaseStudy}, morbidity and mortality rates associated with opioid misuse continue to worsen \cite{overdoseDeaths}. Traditionally, those suffering from addiction take to support groups, such as Alcoholics Anonymous (AA) and Narcotics Anonymous (NA), on their road to recovery \cite{selfHelp}. These groups, which provide an encouraging community and facilitate programs for addiction management and recovery, have shown significant promise in assisting substance abusers \cite{peerSupport}. 
In addition to the aforementioned support groups, federal support is provided through organizations such as the Substance Abuse and Mental Health Services Administration (SAMHSA). Among others, SAMHSA offers programs such as Medication-assisted Treatment and Too Smart To Start - the former combines the use of medication and behavioral therapy in the treatment of substance abuse and the latter is a public education initiative which deters underage alcohol use.

More recently, communities have been established in online social media and discussion forums including Reddit, MedHelp, Twitter and \url{Drugs-Forums.com}, among others. Enacted and internal stigmatization addicts must face \cite{selfStigma,internetUseBerger} in conjunction with the convenience, privacy, and anonymity of such online communities may be explanations for their rapid expansion. These online hubs offer havens for individuals to seek advice, extend support, and share their addiction stories without fear of recourse or judgment. A previous study suggested that these communities can be tremendous resources in the understanding, monitoring and intervening of substance abuse \cite{SCALINGUP}.

Earlier works underscore the utility of mining data from social media sites to understanding trends of human health and drug use \cite{Forum77,TopicModelPaul,Sarker,SuicidePaper,FieldGuidetoLife,eshleman}; however, most of these works do not explore the particular transition from voluntary drug use to compulsive drug use, as well as the various factors that influence such a transition. Our research leverages machine learning and data mining techniques to better understand real-time, unfiltered user data presented to us via Reddit, and in particular, focuses on the understanding of how users transition into addiction can be predicted by using drug utterances and linguistic features contained in their Reddit posts. We present two statistical models: 1) a ``transition classifier" that predicts if a user, given 6 months of content history in general drug discussion forums, will post in forums dedicated to substance recovery support in the subsequent 12 months and 2) a proportional hazards survival model which estimates a user's probability of posting in a recovery forum within the next year. 

\begin{figure*}[t]
    \centering
    \begin{subfigure}[t]{0.22\textwidth}
        \centering
        \includegraphics[height=1.7in]{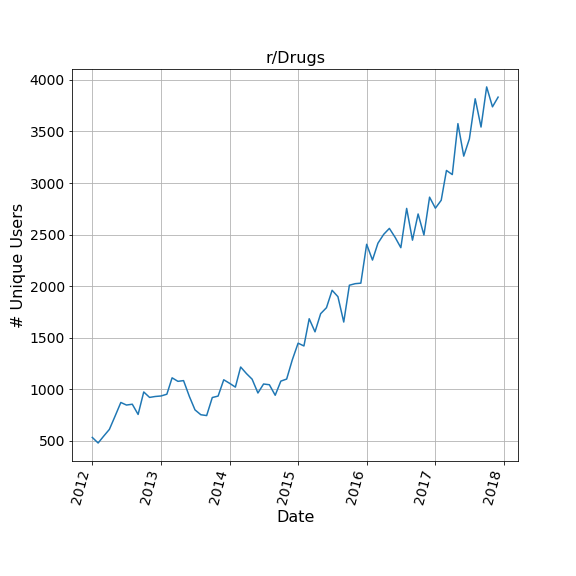}
    \end{subfigure}%
    ~ 
    \begin{subfigure}[t]{0.22\textwidth}
        \centering
        \includegraphics[height=1.7in]{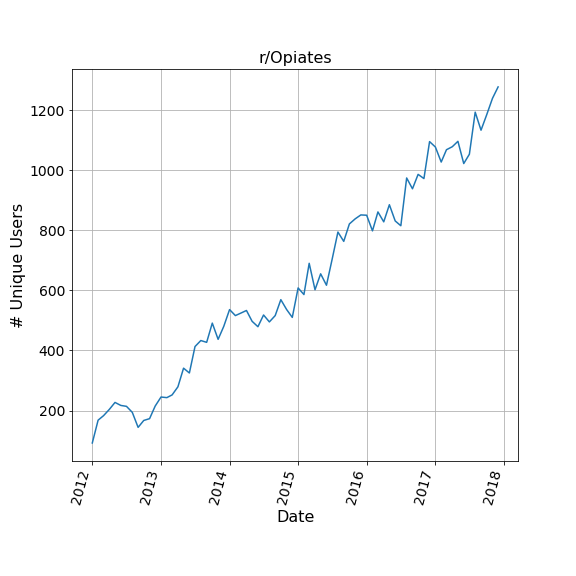}
    \end{subfigure}
     ~ 
    \begin{subfigure}[t]{0.22\textwidth}
        \centering
        \includegraphics[height=1.7in]{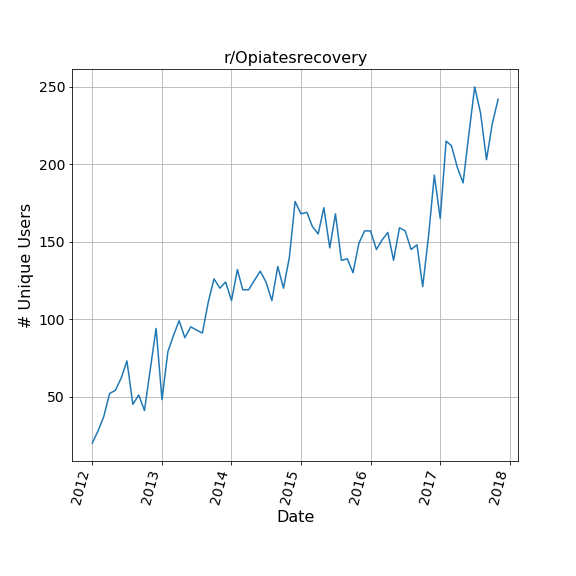}
    \end{subfigure}
    ~ 
    \begin{subfigure}[t]{0.22\textwidth}
        \centering
        \includegraphics[height=1.7in]{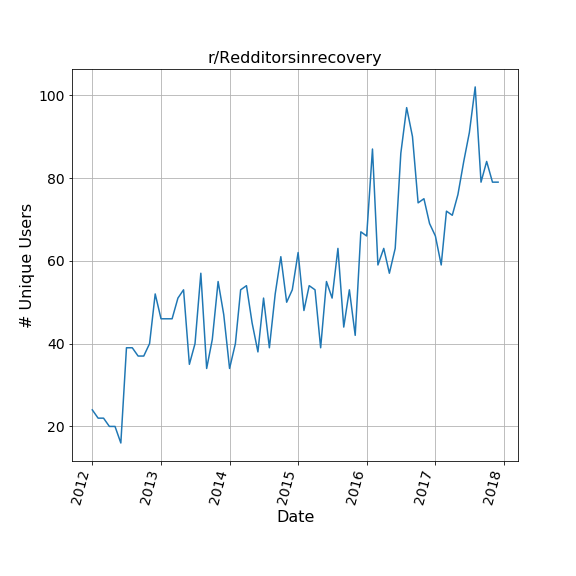}
    \end{subfigure}
   
    \medskip
    \centering
    \begin{tabular}{l|cccc}
    \toprule
    \multicolumn{1}{r}{} & \multicolumn{2}{c}{\textbf{Unique Users}} & \multicolumn{2}{c}{\textbf{User Post Volume}} \\
    \midrule
    \textbf{Subreddit} & \multicolumn{1}{p{9.915em}}{ \% Growth 2012 to 2017} & CAGR  & \multicolumn{1}{p{11.5em}}{\% Growth 2012 to 2017} & CAGR \\
    \midrule
    Drugs & 324\% & 34\%  & 400\% & 38\% \\
    Opiates & 464\% & 41\%  & 657\% & 50\% \\
    RedditorsInRecovery & 387\% & 37\%  & 452\% & 41\% \\
    OpiatesRecovery & 156\% & 21\%  & 180\% & 23\% \\
    \bottomrule
    \end{tabular}%
    \caption{Subreddit Growth, 2012 to 2017}
    \label{fig:Subreddit Growth}
\end{figure*}

\section*{BACKGROUND AND RELATED WORKS}
\subsection*{Drug Addiction and the Opioid Crisis}
Drug addiction is characterized by the compulsive, continued use of a substance despite its negative effects. While drug consumption often begins as recreational, frequent use increases one's tolerance, ultimately altering brain chemistry, inducing heightened desires for drugs, and prompting involuntary and compulsive use. Opioids readily increase tolerance; with each use, one requires larger doses to reach the same level of efficacy. Consequently, opioids are highly addictive, contributing to the high prevalence of opioid misuse \cite{presOpioidWilson}.

Since the early 2000s, the rate of opioid use has climbed, in large part, due to over-prescribing opioid pain relievers such as Oxycodone and Hydromorphone \cite{Kolodny}. A pattern of over-prescribing and related overdose saw its peak in 2011 \cite{PainMedicine} when the Centers for Disease Control and Prevention (CDC) and the Drug Enforcement Administration (DEA) began to address this issue by implementing efforts to educate both medical professionals and the public on appropriate opiate use \cite{presOpioidWilson}. These efforts included tightening prescriptions and developing new prescription opioids that are ``abuse-deterrent" \cite{presOpioidWilson}. While these efforts were met with declines in prescription opioid abuse, illicit opioid use---including that of Heroin and Fentanyl---continued to increase and contribute to rising opioid-related injuries and overdoses \cite{PainMedicine}. One explanation for the continued rise in illicit opioids was supposedly that increased barriers to medication prompted chronic pain patients to supplement their reduced prescription allowance via self-medication, further exacerbating opioid-use risk \cite{PainMedicine}. Since Heroin is pharmacologically similar to prescription opioids, relatively cheap, and readily available, it was an obvious replacement for those previously using prescription opioids \cite{Kolodny}. The trends outlined here demonstrate the need for balanced prevention measures that aim to reduce opioid abuse and overdose while simultaneously maintaining access to prescription opioids and treatment programs as needed  \cite{presOpioidWilson}.

\subsection*{Using Social Media to Understand Drug-Related Issues}
Mining social media data to study health and drug-related behavior is not a new concept. MacLean et al. created a model from MedHelp forums that attempts to predict addiction relapse \cite{Forum77}. Paul and Dredze used a factorial LDA topic model on forums from \url{Drugs-Forums.com} to model drug types, delivery methods, and other related aspects such as cultural and health factors in the context of recreational drug use \cite{TopicModelPaul}. 
Sarker et al. assessed the use of Twitter in analyzing patterns of drug abuse, and built a binary classifier that distinguishes whether or not a tweet contains signs of medication abuse \cite{Sarker}. Other studies have applied machine learning to classify users of an addiction recovery mobile application \cite{FieldGuidetoLife} and to identify distinct behavioral markers between Heroin and amphetamine dependence \cite{BehaviorMarkers}. In the context of Reddit, Choudhury et al. used Reddit data to explore the transition between mental health discourse and suicidal ideation, and built a classifier that distinguishes between those two states \cite{SuicidePaper}. Eshleman et al. demonstrated the possibility of using a binary classifier in predicting a user's propensity to post in a recovery-related subreddit \cite{eshleman}. Our research expands upon this latter work by exploring various linguistic factors and drug utterances in a user's post that are predictive of transitions into substance abuse; further, we build a survival model capable of estimating the probability of such transitions, providing deeper insight into the explanatory factors involved in such shifts.

\section*{REDDIT DATA}
\url{Reddit.com} (or Reddit) is an online collection of threads grouped by user communities known as ``subreddits" with each covering a distinct topic. Reddit users, or ``Redditors," subscribe and submit content to subreddits which interest them, and have their submitted content voted and commented on by fellow Redditors. Reddit has the added appeal of anonymity, allowing users to partake in unfiltered conversations on topics of shared interest. As of April 2018, the platform has over 330 million active users with 130 thousand active subreddits \cite{RedditStats}.

Threads on Reddit are defined by a user's initial post and the subsequent comments on the post by other users. Posts typically discuss a user's own substance use/abuse while comments primarily answer questions asked in the post and/or offer support to the post author. Since our objective is to learn about transitions into substance abuse, which requires analysis of content pertaining to a user's own situation, we restrict our analysis solely to posts.

Using the pushift.io Reddit API\footnote{https://pushshift.io/}, we pulled data from January 2012 through May 2018. This dataset consisted of 309,528 posts from 125,194 unique users and included various attributes of each post such as content, title, author, date of post, number of comments, and number of upvotes.

We focus on four major drug-related subreddits: r/Opiates, r/Drugs, r/OpiatesRecovery, r/RedditorsInRecovery. In addition to having adequate user and post volume, these subreddits host discussions on a variety of drugs. Whereas r/Opiates and r/Drugs primarily serve as forums for general drug discussion, which tend to be more casual in nature, the r/OpiatesRecovery and r/RedditorsInRecovery subreddits provide an avenue for those struggling with substance abuse and addiction to seek advice, share success and relapse stories, and support others. The measurable growth in both user base and post volume between 2012 and 2017 is exhibited in Figure~\ref{fig:Subreddit Growth}. In the proceeding analyses, we will refer to r/Opiates and r/Drugs as ``casual" subreddits and r/OpiatesRecovery and r/RedditorsInRecovery as ``recovery" subreddits.

\section*{Transition Classification: Modelling Transitions from Recreational Use to Substance Abuse}
We trained a binary classifier to model whether a user who posts only in casual subreddits in their first 6 months will eventually go on to post in recovery subreddits in the following 12 months.

\subsection*{Creating Classes}
\begin{figure}[h]
    \centering
    \includegraphics[width=3.4in, height=2.0in]{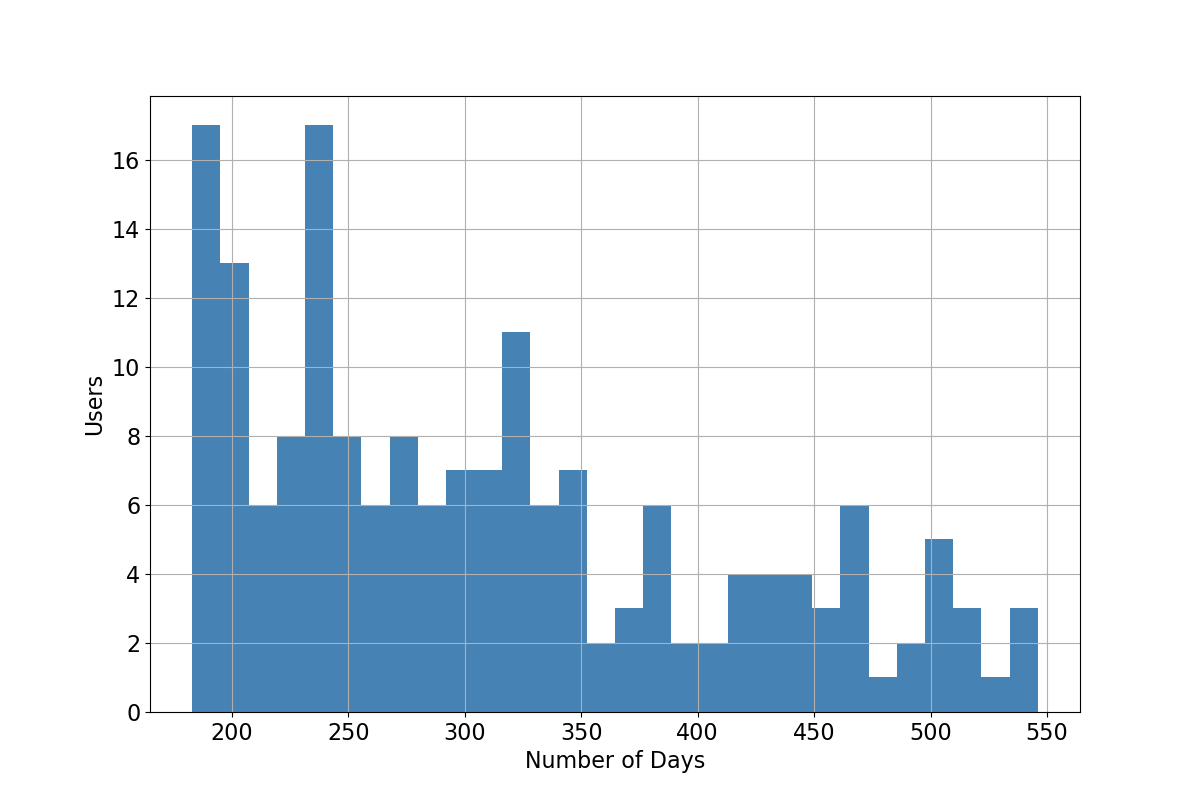}
    \caption{Days Until First Recovery Post in the RECOV Group.}
    \label{fig:Days Until Recov}
\end{figure}

Analysis was restricted to users with at least 3 posts and who exclusively posted in casual subreddits in their first 6 months on Reddit. Of these, we found the subset of users that posted in a recovery subreddit within the next 12 months; there are 220 such users, and we label this group collectively as the CAS$\rightarrow$RECOV class. Figure~\ref{fig:Days Until Recov} shows the distribution of the number of days until the first recovery post for the CAS$\rightarrow$RECOV group.

There are 2,836 users with at least 18 months of casual-only posts. To maintain balance between the two classes, we randomly sample 220 users from this set to form the CAS class. We then split these 440 examples into a training set (352 users) and test set (88 users). 

\subsection*{Feature Selection}
In applying a machine learning model for predicting users' transition from CAS$\rightarrow$RECOV, each user should be represented by a vector. For this purpose, we extract various features from a user's posts,
including dense embedding vectors obtained through $Doc2Vec$, linguistic characteristics of user posts, and drug utterances in the
post. For the last two kinds of features, we use Kruskal-Wallis statistical test to determine which drug utterances and linguistic measures are significantly different between the CAS and CAS$\rightarrow$RECOV classes. Tables~\ref{tab:transition_model_features} and~\ref{tab:transition_model_liwc} show the selected features and their frequency distribution.

\subsubsection*{Doc2Vec Embeddings}
Text clustering and language processing applications involve algorithms that generally require the input text data to be represented as a fix-length vector. Bag-of-word models and n-gram models are often utilized to generate these vector representations due to their simplicity, but both lack contextual information. We therefore use Gensim's Doc2Vec model to create 100-dimensional vector representations for each user post \cite{gensim}. The idea behind such representations was first proposed by Le and Mikolov~\cite{doc2vec}, who sought to create a methodology that generates vector embeddings for texts of variable length. This unsupervised framework has been shown to outperform the bag-of-words and N-gram models in sentiment analysis and other information processing tasks.

Since predictions are based on 6 months of user posts, we aggregate the document vectors of a user's posts over this period. Let $\mathcal{D}_j$ be the set of posts made by user $j$ in the first 6 months on Reddit (ordered by date). Then, denoting the $i$th post of user $j$ by $d^{j}_{i}$, we use the \textit{centroid} of the doc2vec  vectors of user $j$'s posts as a representation of the user $j$.
\begin{align*}
C_j = \frac{1}{|\mathcal{D}_j|}\sum_{i=1}^{|\mathcal{D}_j|} \mathbf{d}^{j}_{i} \ \in \mathbb{R}^{100}
\end{align*}
In the above equation,
$\mathbf{d}_i^j$ is the doc2vec based vector representation of $d_i^j$.
 Though simplistic, prior works \cite{centroidRadev,docSumBOW} have suggested the efficacy of using the centroid of document representations to capture meaningful content from a set of documents.

\subsubsection*{Linguistic Measures}
The specific language and words one employs in speech and writing can reveal much about one's psychological and social states \cite{pennebaker_2013, pennebaker_mehl_niederhoffer_2003}. Further, studies have shown socio-psychological and personality differences between drug and non-drug addicts \cite{personality_differences}. To capture both linguistic components (e.g. fraction of pronouns, verbs, and articles among others) and psychological aspects (e.g. words associated with positive/negative emotions, anger, and family/friends among others) of user posts, we use the Linguistic Inquiry and Word Count\footnote{Language Inquiry and Word Count: http://liwc.wpengine.com/} (LIWC), a text analysis program that categorizes words into 93 groups which reflect different emotions, thinking styles, social concerns, and parts of speech. Though LIWC variables capture more than purely linguistic dimensions of texts, we will use the phrases ``LIWC features" and ``linguistic features" interchangeably hereafter for simplicity.

\subsubsection*{Keywords}
We used odds ratios -- a metric used in statistics to measure the  association between the presence of one property with the presence of another -- to find discriminative keywords for each class such as (e.g. ``experiences" for CAS and ``addiction" for CAS$\rightarrow$RECOV). More concretely, if $\mathcal{W}$ is the set of all words in our training data, then the odds ratio, $OR(c,w)$, for a word, $w \in \mathcal{W}$ and class $c \in \{CAS, CAS \rightarrow RECOV \}$ is given by,

\begin{align*}
OR(c,w) = \frac{\frac{freq(c,w)}{freq(\neg{c},w)}}{\frac{\neg{freq(c,w)}}{\neg{freq(\neg{c},w)}}}
= \frac{freq(c,w)*\neg{freq(\neg{c},w)}}{freq(\neg{c},w)*\neg{freq(c,w)}}
\end{align*}
where $freq(c,w)$ is the number of posts in class $c$ in which the word $w$ occurs and $\neg{freq(c,w)}$ is the number of posts in class $c$ where word $w$ does not occur. The odds ratio quantifies how strongly associated $w$ is with the class $c$; a higher odds ratio implies a stronger association of $w$ with $c$.

We choose $w$ to be a keyword for $c$ if
$OR(c,w)>2$ and $|OR(c,w)-OR(\neg{c},w)|>2$. That is, $w$ has a high OR with respect to one class and a substantially lower $OR$ with respect to the other class. We list a sample of associated keywords for both classes in Table~\ref{tab:class_keywords}.

\begin{table}[htbp]
  \centering
  \caption{Discriminatory keywords for CAS and CAS$\rightarrow$RECOV class using Odds Ratio}
    \begin{tabular}{c|p{20.915em}}
    \multicolumn{1}{c}{\textbf{Class}} & \multicolumn{1}{c}{\textbf{Keywords}} \\
    \midrule
    CAS   & friends, completely, lsd, trip, reddit, music, weird, dr, experiences, friend, ended, 100, mdma \\
    \midrule
    CAS$\rightarrow$RECOV & quit, addiction, clean, anymore, subs, suboxone, oxy, bupe \\
    \end{tabular}%
  \label{tab:class_keywords}%
\end{table}%

\subsubsection*{Drug Utterances}
\begin{figure}[h]
    \centering
    \caption{Drugs with statistically significant variation in utterances between CAS and CAS$\rightarrow$RECOV (p-values $<$ 0.05 using Kruskal-Wallis test).}
    \includegraphics[height=2.4in]{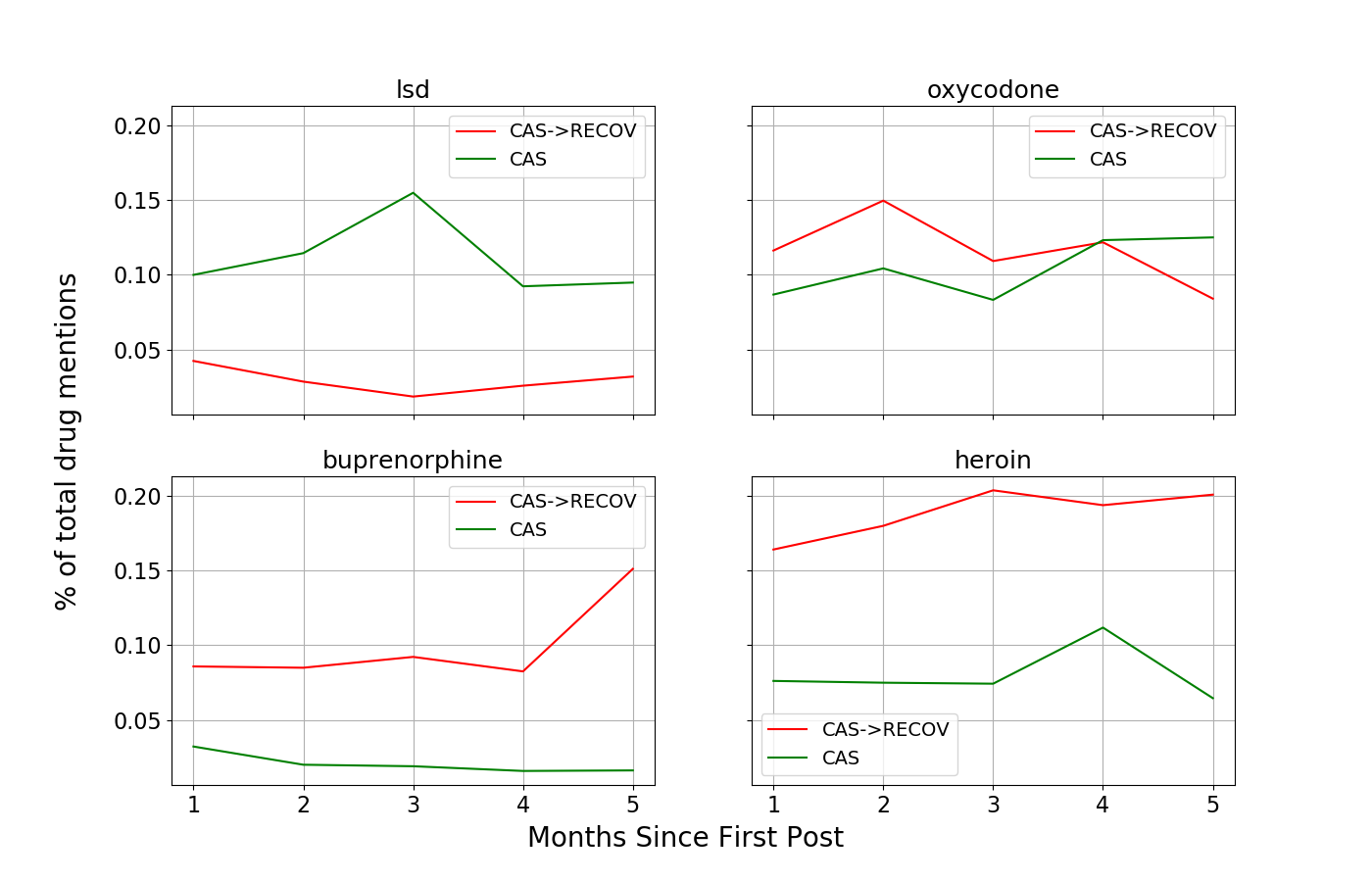}
    \label{fig:drug_mentions}
\end{figure}

Drug utterances serve as an indicator of which drugs users are interested in and/or currently using. For each user we calculate the mentions of each drug (as a \% of all drugs they mention). This calculation includes both formal drug names of the drug, colloquial (``street") names, and major brand names (e.g., OxyContin\textsuperscript{\textregistered} is a common brand of oxycodone and ``coke" is a oft-used name for cocaine). 
In Figure~\ref{fig:drug_mentions}, we displays four drugs that have highly significant variation in utterance between the two classes.

\section*{Survival Analysis: Predicting When Transitions are Likely to Occur}
For our task, a traditional classification model predicts whether
a transition would occur from CAS to CAS$\rightarrow$RECOV. However, for such information to be useful in real-life, it is
also important to know \textit{when} such transitions might occur.
This additional information can help prioritize vulnerable individuals
based on their propensity to be victim of drug addiction.
In this section, we present a Cox (proportional hazards) regression survival model to solve this task. This model also enables us to investigate the effect of different variables on the predicted time until one's first recovery post.

The general formulation of a survival model centers around a random variable, $T$ (denoting time), the survival function, 
\begin{equation*}
S(t) = Pr(T > t)
\end{equation*}
and the hazard function,
\begin{equation*}
\lambda(t) = \frac{f(t)}{S(t)} \quad [where \ f(t) = \frac{d}{dt}(1-S(t))]
\end{equation*}
In our case, $S(t)$ is the probability that a user will last (``survive") more than $t$ days without posting in a recovery subreddit and $\lambda(t)$ is the instantaneous rate of transitions to recovery subreddits (among those users that have never posted in a recovery subreddit) at time $t$. Adopting terminology commonly used in survival analyses, we shall say a user ``survives" if he lasts more than 12 months without a recovery post and ``fails" otherwise.

\subsection*{Right-Censored Data}
Besides providing the answer whether a user will survive 
beyond a given time, survival model has another crucial advantage: a survival model can utilize ``censored data'' effectively, whereas a traditional regression model fails to do so. Censored data (also known as right-censored data) refers to observations for which the desired event
has not happened yet. For instance, say, the end of observation time for our study is $\hat{t}$. Now, if a user has never posted in recovery forum within our observation period, one does not know with certainty whether that user will not post in recovery on some day $t > \hat{t}$. This is an example of \textit{right-censored} data 
instances.

Our model requires users to have at least 10 posts where the first 3 were casual and not all occurring on the same day. This criteria eliminates the case of 0-day survival times, which in real-life makes little sense. In our dataset, there are 2,367 users satisfying these restrictions, and for each user, we look only at (up to) their first 12 months of posts. 165 of these users fail within the first 12 months, while the remainder are right-censored observations. 

\subsection*{Cox Regression}
A Cox regression model is a specific type of survival model that accounts for the effects of covariates on some baseline hazard function $\lambda_0(t)$. Formally, if we let $\bm{x^{(i)}} \in \mathbb{R}^d$ be a column vector of features for user $i$, the \textit{hazard} for user $i$ is,
\begin{equation*}
\lambda_i(t) = \lambda_0(t)\exp\{\bm{x}^\intercal \bm{\beta^{(i)}}\}
\end{equation*}
and the corresponding survival function is,
\begin{equation*}
S_i(t) = S_0(t)^{\exp\{\bm{\beta}^\intercal \bm{x^{(i)}}\}}
\end{equation*}
where $\bm{\beta} \in \mathbb{R}^d$ is a vector of trainable parameters. The data for the model can be denoted $D_{surv} = \{(y_i, \delta_i, \bm{x}^{(i)}) : i = 1, 2...,n\}$ where $y_i$ is the minimum of the censoring time $C_i$ (end of observation time) and survival time $T_i$ and 
\[\delta_i =
	\begin{cases} 
      1 & \ if \ \ T_i = y_i \\
      0 & \ otherwise
   \end{cases},
\]
where $\delta_i$ denotes whether or not an instance is censored.
We train $\bm{\beta}$ by maximizing the partial likelihood estimate (under iid assumption),
\begin{equation*}
L(\bm{\beta}|D_{surv}) = \mathlarger{\prod}_{i=1}^{n} \left[\frac{\lambda_0(t) \exp\{\bm{\beta}^\intercal \bm{x^{(i)}}\}}{\sum_{j \in \Psi(y_i)}\lambda_0(t) \exp\{\bm{\beta}^\intercal \bm{x^{(j)}}\}}\right]^{\delta_i}
\end{equation*}
where $\Psi(t)=\{i:y_i>t\}$ is the subset of users who survive past time $t$.
\section*{RESULTS}
In this section we provide experimental results for both binary classification and cox regression. Results for both models suggest that drug utterances and linguistic features can be indicative of one's propensity to shift towards substance abuse.

We use categorical accuracy and F1 score to evaluate the transition classifier. A baseline model, which used only Doc2Vec embeddings, achieved a modest $69.3\%$ test-set accuracy. With the inclusion of LIWC linguistic features, drug utterances, and class-specific keywords in addition to tuning model parameters through grid search, we improved accuracy by roughly $5\%$ (Table~\ref{tab:transition_classifier_results}).
\begin{table}[htbp]
  \centering
  \caption{Transition Classifier Results Summary. \small Table displays test-set performance of Random Forest with 170 trees (selected using grid search and 10-fold cross validation) using different features. Model trained on users' first 6 months of posts and predicts transitions in the subsequent 12 months. Number of train and test set examples were 352 and 88, respectively.}
    \begin{tabular}{l|cc}
    \toprule
    \multicolumn{1}{c}{Model} & Accuracy & F1 Score \\
    \midrule
    Doc2Vec & 0.693 & 0.682 \\
    LIWC  & 0.659 & 0.659 \\
    Doc2Vec + drugs + keywords & 0.716 & 0.725 \\
    LIWC + drugs + keywords & 0.739 & 0.736 \\
    \textbf{Doc2Vec + LIWC + drugs + keywords} & \textbf{0.750} & \textbf{0.750} \\
    \bottomrule
    \end{tabular}%
  \label{tab:transition_classifier_results}%
\end{table}

Performance of the survival model was evaluated using Concordance Index (C-Index), a standard metric often employed in survival models \cite{c-index}. Concordance ranges from 0 to 1 (a higher score indicates a stronger model) and is analagous to the AUC score of the Receiver Operating Characteristic (ROC). Our best Cox model achieved a 0.823 C-Index on the test-set (Table~\ref{tab:survival_results}), indicating a moderately strong model.
\begin{table}[htbp]
	\centering
	\caption{Cox Model Results Summary. \small Train/test split of 1,775 (1665 censored) and 592 (352 censored) users, respectively. C-Index shown for models using different feature sets. The model using drug utterances, keywords, and LIWC features performed best on training set using 5-fold cross validation and gave a test-set C-Index of 0.820. Test set data consisted of 45 observed and 592 censored examples.}
	\begin{tabular}[t]{l|c}
		\toprule
		\multicolumn{1}{c}{Model} & C-Index \\
		\midrule
		Doc2Vec  & 0.790 \\
		Doc2Vec + drugs + keywords + LIWC & 0.788 \\
		\textbf{Drugs + keywords + LIWC} & \textbf{0.820} \\
		\midrule
		\textit{\textbf{Test Set Performance}} & \textit{\textbf{0.820}} \\
		\bottomrule
	\end{tabular}%
	\label{tab:survival_results}
\end{table}%

In addition to training Cox models using different feature sets, we sought to explore the explanatory strength of individual covariates by fitting a model to each individual feature (Table~\ref{tab:cox_individual} presents the C-statistics from this experiment). Using this approach, we found that utterances of drugs such as Buprenorphine, Heroin, and LSD, have a stronger impact on a user's predicted survival probability relative to other drugs. This is in accordance with our earlier analysis (Figure~\ref{fig:drug_mentions}). Similarly, LIWC dimensions such as ``leisure," ``time," and ``focuspresent" have relatively more predictive power (Table \ref{tab:cox_individual}). These features measure the extent to which posts contain terms related to leisure activities (e.g. ``cook," ``chat," ``movie"), time-related terms (e.g. ``hour," ``day," ``oclock"), and terms that indicate a focus on the present (e.g ``today," ``is," ``now"). 
\begin{table}[htbp]
  \centering
  \caption{Top 10 Explanatory Covariates}
  \label{tab:cox_individual}
  \begin{subtable}{0.23\textwidth}
    \begin{tabular}{lc}
    \toprule
    Drug Name & C-Index \\
    \midrule
    Heroin & 0.748 \\
    Buprenorphine & 0.702 \\
    LSD   & 0.687 \\
    psilocybin & 0.628 \\
    oxycodone & 0.623 \\
    marijuana & 0.621 \\
    Ecstasy & 0.614 \\
    fentanyl & 0.610 \\
    oxymorphone & 0.608 \\
    amphetamine & 0.597 \\
    \bottomrule
    \end{tabular}%
  \end{subtable}%
  \hspace{\fill}
  \begin{subtable}{0.23\textwidth}
    \begin{tabular}{lc}
    \toprule
    LIWC feature & C-Index \\
    \midrule
    leisure & 0.668 \\
    Period & 0.646 \\
    time  & 0.646 \\
    ingest & 0.645 \\
    informal & 0.642 \\
    netspeak & 0.633 \\
    focuspresent & 0.630 \\
    relativ & 0.627 \\
    nonflu & 0.612 \\
    money & 0.610 \\
    \bottomrule
    \end{tabular}%
  \end{subtable}%
\end{table}%

\subsection*{Survival Predictions on the Transition Dataset}
\begin{figure}[h]
  \centering
  \caption{Surviving One Year. \small Histograms showing the number of CAS and CAS$\rightarrow$RECOV users predicted to survive at least a year.}
  \label{fig:survival differences}
  \includegraphics[height=1.3in, keepaspectratio]{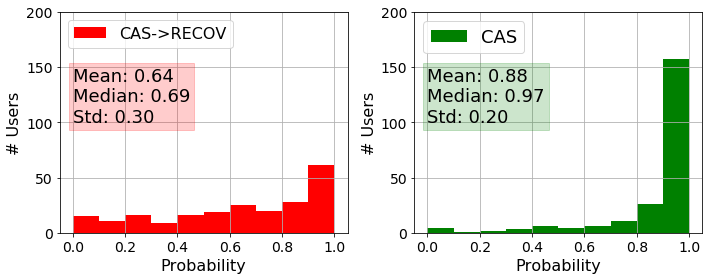}
\end{figure}

We looked once again at the 220 CAS and 220 CAS$\rightarrow$RECOV Redditors, this time through the lens of our trained survival model. In Figure~\ref{fig:survival differences}, we fix the time duration at 12 months and compare survival probabilities between the two groups. Not surprisingly, a sizable majority of the CAS group have high probability ($>90\%$) of surviving past 12 months.
We then approximated the addictive potential of certain drugs by measuring the average survival probability of users who share a common drug of choice. For example, we found that users whose top drug utterances were Ecstasy or LSD had high probability of surviving while users whose drug of choice is Heroin or Buprenorphine had a comparatively smaller chance of surviving (Table~\ref{tab:Drug Addictive Potential}).
\begin{table}[htbp]
  \centering
  \caption{One-Year Survival Probability by Top Drug Mention}
  \begin{tabular}{lclc}
  \toprule
  \textbf{Drug Name} & \multicolumn{1}{p{5.335em}}{\textbf{ Surv. Prob.}} & \textbf{Drug Name} & \multicolumn{1}{p{5.665em}}{\textbf{ Surv. Prob.}} \\
  \midrule
  Ecstasy & 0.987 & fentanyl & 0.820 \\
  LSD   & 0.981 & cocaine & 0.774 \\
  benzodiazepines & 0.877 & oxycodone & 0.767 \\
  marijuana & 0.872 & Heroin & 0.502 \\
  methamphetamine & 0.824 & Buprenorphine & 0.498 \\
  \bottomrule
  \end{tabular}%
  \label{tab:Drug Addictive Potential}%
\end{table}%

\section*{Case Study}
In this section, we illustrate one of several potential uses of our models by applying them to an actual Redditor. Figure 6 presents a profile of the user including a redacted sample of his writing, the most prevalent LIWC aspects, and most uttered drugs from his posts. 

The subject is a routine user of Heroin and an active participant in r/Opiates. His top 3 drug utterances are Heroin, oxycodone, and Buprenorphine with Heroin representing approximately 35\% of his total drug mentions. Referring to Table~\ref{tab:Drug Addictive Potential}, one may expect a lower 12-month survival probability for this user relative to someone with a random drug composition. Furthermore, LIWC dimensions of the subject's posts are consistent with Redditors who do not survive 1 year. For example, users who fail tend to have a lower ``focuspresent" dimension, and our subject scores close to the $75th$-percentile in this category (Table~\ref{tab:Survival Model LIWC}).


\subsection*{Model Predictions of Subject}
\begin{figure}[h]
  \centering
  \includegraphics[width=3.2in]{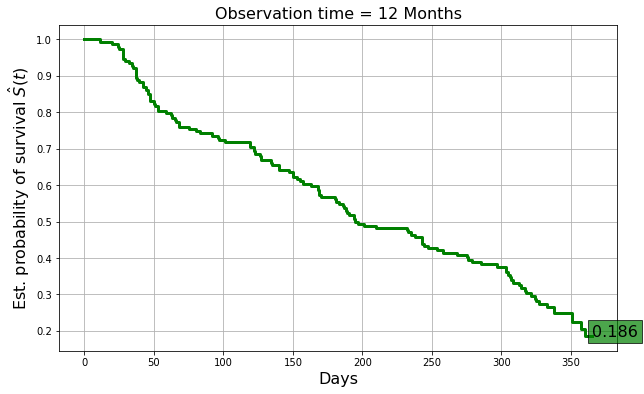}
  \caption{Kaplan-Meier Curve showing surviving probability vs Days for the case study subject}
  \label{fig:Case Study Kaplan-Meier}
\end{figure}

Given the subject's background and profile, it is encouraging that our transition classifier labels him as CAS$\rightarrow$RECOV. That is, using only the subject's first 6 months of casual posts, the classifier predicts he will post in recovery within the subsequent 12 months. Furthermore, our Cox model predicts a less than a 18.6\% chance he will survive past 1 year (Figure~\ref{fig:Case Study Kaplan-Meier}), indicative of a high-risk user. Consulting the subject's entire post history, we found that he did eventually post in r/OpiatesRecovery 200 days after his first post.

\begin{figure*}[t]
  \centering
  
  \label{fig:Case Study}
  \begin{subtable}[t]{.50\textwidth}
  	\centering
    \begin{tabular}[t]{|p{26.5em}|}
    \toprule
    \multicolumn{1}{|c|}{\textbf{Initial Posts}} \\
    \midrule
    Finished my stash off this morning as was expecting to have more by now. Just had a call..... Its not good. Hopefully tomorrow he said. HOPEFULLY..... Tonight is going to be a very long night for me. \\
    \midrule
    ...Ran out of \textbf{dope} last night, meaning to get more this morning. My man was out so planned on getting it while at work. But, I forgot my cash. DAMN, I have a whole night shift in front of me while starting to get sick. This sucks. Hope your all having a better night than me. \\
    \midrule
    ...I have aquired 15 x 100mcg \textbf{Fent} patches of the matrix type. I figured I would try smoking a bit of one and...I now know why its so easy to OD on them....Gonna be super careful with them though, lost 2 good friends over the years through \textbf{fent} OD's. Will keep posting at least once a day until their gone though.  Happy nods ppl, and keep safe. \\
    \bottomrule
    \end{tabular}%
  \end{subtable}%
  \begin{subtable}[t]{.50\textwidth}
  	\centering
    \begin{tabular}[t]{|p{15.5em}|}
    \toprule
    \multicolumn{1}{|c|}{\textbf{First Recovery Post}} \\
    \midrule
    Hi guys, I figured this would be as good a place as any to post my story, and look for answers. I have been using \textbf{Heroin} on and off for about some years now...I am past the physical withdrawals now, but still really struggling with the psychological side...All my friends use, hell, most people I know use. I want to stop for my wife and son above everything else, bit finding it really hard. So, what advice do you guys have. Any will be greatly appreciated. I'm guessing I just want someone to talk to as today I am really struggling with it. Thanks \\
    \bottomrule
    \end{tabular}
  \end{subtable}
  \bigskip
  \begin{subtable}{0.5\textwidth}
    \centering
    \begin{tabular}{lclc}
    \toprule
    LIWC  & Avg. Value & LIWC  & Avg. Value \\
    \midrule
    Authentic & 71.261 & netspeak & 2.176 \\
    relativ & 16.059 & percept & 2.028 \\
    focuspresent & 13.495 & leisure & 1.189 \\
    Period & 11.902 & money & 0.862 \\
    WPS   & 10.648 & family & 0.629 \\
    Sixltr & 9.534 & nonflu & 0.579 \\
    time  & 8.281 & sad   & 0.344 \\
    conj  & 5.841 & death & 0.234 \\
    informal & 4.130 & ingest & 0.215 \\
    insight & 2.491 & Exclam & 0.000 \\
    \bottomrule
    \end{tabular}%
    \vspace{1em}
    \caption{Subset of LIWC}
    \label{tab:Case Study LIWC}
  \end{subtable}%
  \begin{subfigure}{0.5\textwidth}
      \centering
      \includegraphics[height=2.2in]{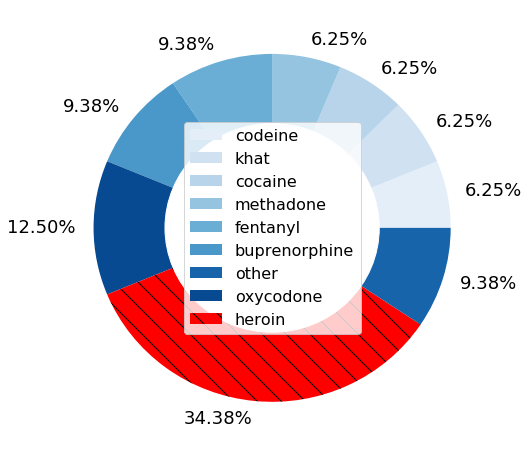}
      \caption{Top Drugs}
      \label{fig:Case Study Top Drugs Pie}
  \end{subfigure}%
  \vspace{-0.2in}
  \caption{Case Study Subject Profile}
  
\end{figure*}%

\section*{DISCUSSION}
In this work, we explored transitions from recreational substance use to substance abuse using survival analysis and binary classification. In our classifier, we were able to predict whether a previously “casual” user will post in a recovery subreddit within the next 12 months. Using our survival model, we were able to uncover distinct features such as specific drug utterances and linguistic features of a given post that influence a given user's probability of transitioning from recreational substance use to substance abuse. In this section, we discuss the implications of these results in informing drug use culture and possible methods of drug classifications, as well as highlight potential applications of our models in understanding drug use from the user's perspective. 

\subsection*{Informing Culture and Classifications}
Based on our Redditors' posts, a pattern of use was apparent: drugs such as LSD, Ecstasy, marijuana, and alcohol were often used by Redditors attending parties, music festivals, or social events in general. In contrast, opioids were commonly used as a habitual, lone activity or in smaller, low-energy gatherings. This trend learned by inspection was further supported by post-level analyses showing a co-occurrence of frequent mentions of LSD and Ecstasy and frequent use of words such as ``music,'' ``rave, '' and ``party.'' Paul and Dredze found similar results in their use of topic modeling to understand various factors, including the cultural factors, associated with certain drug use. For example, the ``culture" words associated with Ecstasy include [``music", ``people", ``great", ``rave"], while the ``culture" words associated with opioids include [``life", ``years", ``money", ``time", ``shit"] \cite{TopicModelPaul}.

Our survival model suggests that certain drug utterances were highly explanatory in influencing the probability of substance abuse -- for instance, a user with higher Heroin mentions (as a percentage of total drug mentions) is associated with a higher probability of substance abuse, while higher LSD and Ecstasy mentions are associated with lower probabilities. This particular example directly conflicts with the current national drug classifications outlined by the DEA, who classifies drugs into five ``schedules'' based on their dependence potential as well as the level of accepted medical use. As it currently stands, drugs including Heroin, LSD, marijuana, and Ecstasy are classified as Schedule I, while most other opioids along with cocaine are classified as Schedule II. Our models, however, suggest that the drugs of choice for individuals who are likely to transition are often not the drugs labeled as Schedule I, with the exception of Heroin. Table \ref{tab:Drug Addictive Potential} outlines the average ``survival" probabilities associated with drugs common among our users. In our model, users whose drugs of choice are Ecstasy, LSD, and marijuana have a higher predicted survival rate which may suggest that these drugs could be better classified as Schedule II or lower. Similarly, their lower associated survival probabilities may be justification for a higher classification for drugs like Buprenorphine, oxycodone, and cocaine.

Other studies have reached similar conclusions.
The United Kingdom has similar drug classifications based on addictive potential, and places many of the Schedule I drugs as their top dangerous drugs as well. Yet Nutt and colleagues ~\cite{UKpaper} found that the UK system of classification is somewhat arbitrary and not driven by scientific evidence. In their work, they reclassified the drugs listed in the UK's guidelines using potential of harms of individual drugs, and found that the top drugs with high potential of harm include Heroin, cocaine, barbiturates, street methadone, and ketamine. Their findings also disagree with UK's designation of LSD and Ecstasy in the top drug class -- this aligns with our results and further suggests the higher potential of harm from opioids relative to these other drugs.

Similarly, Sarvet et al~\cite{medicalmarijuana} discuss the increasing amount of evidence in support of the therapeutic properties of medical marijuana. According to them, medical experts, including the American Medical Association, have urged the DEA to reschedule marijuana from Schedule I to Schedule II. They argue that not only would it increase access for patients who could benefit from this form of treatment, it would also enable further research and development of cannabinoid-based medicine. 

\subsection*{Limitations}
In this section, we list some limitations to our work:
1) We identify selection bias in our Redditors, as they are choosing to express their opinions on Reddit and are likely much more open about their drug use/recovery progress. Since this may not be indicative of all drug users, our results may not be generalizable to the larger population. 
2) We use participation in general drug forums and recovery forums as proxies for recreational use and substance abuse, respectively. However, since there are no rules preventing abuse-related posts in our casual subreddits, these subreddits, though predominately about casual drug discussion, do contain abuse-related posts.
3) We cannot make any clinical disgnoses of drug addition or mental illness based solely on our analyses.
4) Although we made attempts to include all forms of drug names in our drug counts, there may be names that are not included, and thus we may not have captured all types of drug utterances. 
5) Our research uncovers certain post-level characteristics that influence the likelihood of transitions into recovery subreddits but cannot explain \textit{why} Redditors make these transitions. 

Despite these limitations, our findings provide unique insight into drug use patterns from the perspective of users and uncovered features within one's posts that may be predictive of future substance abuse. Our work demonstrates the utility of online forums and social media sources in understanding human health and activity. Finally, the computational models that we provide can
be utilized into real-life applications. For example, there can be
a software App for Redditors in which they input their posts and the App returns the poster's propensity for drug abuse. Also,
this App can be used as a predictive device to help counselors and psychologists in advising their patients.

\section*{Future Work}
Further research in this area could focus more on the temporal aspect of a user's post sequence. Similar to the work done by Maclean et al. \cite{Forum77}, which used a Conditional Random Field (CRF) to model phases of addiction, one could construct a sequence model using our subreddits to track users' phases of ``pre-addiction" to further analyze transitions into addiction and uncover more contextual factors that influence such transitions. We hope to explore such analysis in the future. 






\bibliographystyle{ieeetr}

\bibliography{main.bib}

\begin{thebibliography}{10}

\bibitem{painManagement}
R.~J. Bonnie, M.~A. Ford, and J.~K. Phillips, {\em Pain Management and the
  Opioid Epidemic: Balancing Societal and Individual Benefits and Risks of
  Prescription Opioid Use}.
\newblock The National Academies Press, 2017.

\bibitem{OhioCaseStudy}
J.~Penm, N.~J. MacKinnon, J.~M. Boone, A.~Ciaccia, C.~McNamee, and E.~L.
  Winstanley, ``Strategies and policies to address the opioid epidemic: A case
  study of ohio,'' {\em Journal of the American Pharmacists Association},
  vol.~57, no.~2, Supplement, pp.~S148 -- S153, 2017.

\bibitem{overdoseDeaths}
P.~Seth, L.~Scholl, R.~Rudd, and B.~S, ``Overdose deaths involving opioids,
  cocaine, and psychostimulant -- united states, 2015--2016,'' vol.~67,
  pp.~349--358, 2018.

\bibitem{selfHelp}
F.~A. with NCADD, ``Multiple pathways of recovery.''
\newblock
  https://www.facingaddiction.org/resources/multiple-pathways-of-recovery.

\bibitem{peerSupport}
S.~P.~W. Kathlene~Tracy, ``Benefits of peer support groups in the treatment of
  addiction,'' {\em Substance Abuse and Rehabilitation}, vol.~7.

\bibitem{selfStigma}
J.~B. Luoma, B.~S. Kohlenberg, S.~C. Hayes, K.~Bunting, and A.~K. Rye,
  ``Reducing self-stigma in substance abuse through acceptance and commitment
  therapy: Model, manual development, and pilot outcomes,'' {\em Addiction
  Research \& Theory}, vol.~16, no.~2, pp.~149--165, 2008.

\bibitem{internetUseBerger}
M.~Berger, T.~H. Wagner, and L.~C. Baker, ``Internet use and stigmatized
  illness,'' {\em Social Science and Medicine}, vol.~61, no.~8, pp.~1821 --
  1827, 2005.

\bibitem{SCALINGUP}
K.~J. Sunny, M.~A. Lisa, H.~T. Jeffrey, and D.~K. Amarendra, ``Scaling up
  research on drug abuse and addiction through social media big data,'' {\em J
  Med Internet Res}, vol.~19, p.~e353, Oct 2017.

\bibitem{Forum77}
D.~MacLean, S.~Gupta, A.~Lembke, C.~Manning, and J.~Heer, ``Forum77: An
  analysis of an online health forum dedicated to addiction recovery,'' in {\em
  Proc. of the 18th ACM Conference on Computer Supported Cooperative Work \&
  Social Computing}, pp.~1511--1526, 2015.

\bibitem{TopicModelPaul}
M.~J. Paul and M.~Dredze, ``Experimenting with drugs (and topic models):
  Multi-dimensional exploration of recreational drug discussions,'' in {\em
  Proc. of AAAI}, 2012.

\bibitem{Sarker}
A.~Sarker, K.~O'Connor, R.~Ginn, M.~Scotch, K.~Smith, D.~Malone, and
  G.~Gonzalez, ``Social media mining for toxicovigilance: Automatic monitoring
  of prescription medication abuse from twitter,'' {\em Drug Safety}, vol.~39,
  pp.~231--240, Mar 2016.

\bibitem{SuicidePaper}
D.~Choudhury, Munmun, Kiciman, Emre, M.~Dredze, G.~Coppersmith, and M.~Kumar,
  ``Discovering shifts to suicidal ideation from mental health content in
  social media,'' in {\em Proc. of the CHI Conference on Human Factors in
  Computing Systems}, pp.~2098--2110, 2016.

\bibitem{FieldGuidetoLife}
B.~Fischman, ``Data driven support for substance addiction recovery
  communities,'' in {\em Extended Abstracts of the 2018 CHI Conference on Human
  Factors in Computing Systems}, CHI EA '18, (New York, NY, USA),
  pp.~SRC07:1--SRC07:6, ACM, 2018.

\bibitem{eshleman}
R.~Eshleman, D.~Jha, and R.~Singh, ``Identifying individuals amenable to drug
  recovery interventions through computational analysis of addiction content in
  social media,'' {\em 2017 IEEE International Conference on Bioinformatics and
  Biomedicine (BIBM)}, 2017.

\bibitem{presOpioidWilson}
W.~M. Compton, C.~M. Jones, and G.~T. Baldwin, ``Relationship between
  nonmedical prescription-opioid use and heroin use,'' {\em New England Journal
  of Medicine}, vol.~374, no.~2, pp.~154--163, 2016.

\bibitem{Kolodny}
A.~Kolodny, D.~T. Courtwright, C.~S. Hwang, P.~Kreiner, J.~L. Eadie, T.~W.
  Clark, and G.~C. Alexander, ``The prescription opioid and heroin crisis: A
  public health approach to an epidemic of addiction,'' {\em Annual Review of
  Public Health}, vol.~36, no.~1, pp.~559--574, 2015.

\bibitem{PainMedicine}
M.~E. Rose, ``Are prescription opioids driving the opioid crisis? assumptions
  vs facts,'' {\em Pain Medicine}, vol.~19, no.~4, pp.~793--807, 2018.

\bibitem{BehaviorMarkers}
Ahn, Woo-Young, Vassileva, and Jasmin, ``Machine-learning identifies
  substance-specific behavioral markers for opiate and stimulant dependence,''
  {\em Drug and Alcohol Dependence}, vol.~161, 2016/04/01.

\bibitem{RedditStats}
A.~Hutchinson, ``Reddit now has as many users as twitter, and far higher
  engagement rates.''
\newblock
  https://www.socialmediatoday.com/news/reddit-now-has-as-many-users-as-twitter-and-far-higher-engagement-rates/521789/.

\bibitem{gensim}
``Gensim: Doc2vec paragraph embeddings.''
\newblock https://www.radimrehurek.com/gensim/models/doc2vec.html.

\bibitem{doc2vec}
Q.~Le and T.~Mikolov, ``Distributed representations of sentences and
  documents,'' in {\em Proceedings of the 31st International Conference on
  International Conference on Machine Learning - Volume 32}, ICML'14,
  pp.~II--1188--II--1196, JMLR.org, 2014.

\bibitem{centroidRadev}
D.~R. Radev, H.~Jing, M.~Stys, and D.~Tam, ``Centroid-based summarization of
  multiple documents,'' {\em Inf. Process. Manage.}, vol.~40, pp.~919--938,
  Nov. 2004.

\bibitem{docSumBOW}
K.~Mani, I.~Verma, and L.~Dey, ``Multi-document summarization using distributed
  bag-of-words model,'' {\em CoRR}, vol.~abs/1710.02745, 2017.

\bibitem{pennebaker_2013}
J.~W. Pennebaker, {\em The secret life of pronouns: what our words say about
  us}.
\newblock Bloomsbury Press, 2013.

\bibitem{pennebaker_mehl_niederhoffer_2003}
J.~W. Pennebaker, M.~R. Mehl, and K.~G. Niederhoffer, ``Psychological aspects
  of natural language use: Our words, our selves,'' {\em Annual Review of
  Psychology}, vol.~54, no.~1, pp.~547--577, 2003.

\bibitem{personality_differences}
M.~R. Gossop and S.~B. Eysenck, ``A further investigation into the personality
  of drug addicts in treatment,'' {\em Addiction}, vol.~75, no.~3,
  pp.~305--311, 1980.

\bibitem{c-index}
H.~Steck, B.~Krishnapuram, C.~Dehing-oberije, P.~Lambin, and V.~C. Raykar, ``On
  ranking in survival analysis: Bounds on the concordance index,'' in {\em
  Advances in Neural Information Processing Systems 20}, pp.~1209--1216, Curran
  Associates, Inc., 2008.

\bibitem{UKpaper}
D.~Nutt, L.~A. King, W.~Saulsbury, and C.~Blakemore, ``Development of a
  rational scale to assess the harm of drugs of potential misuse,'' {\em The
  Lancet}, vol.~369, no.~9566, pp.~1047 -- 1053, 2007.

\bibitem{medicalmarijuana}
A.~L. Sarvet, M.~M. Wall, D.~S. Fink, E.~Greene, A.~Le, A.~E. Boustead, R.~L.
  Pacula, K.~M. Keyes, M.~Cerdá, S.~Galea, and D.~S. Hasin, ``Medical
  marijuana laws and adolescent marijuana use in the united states: a
  systematic review and meta-analysis,'' {\em Addiction}, vol.~113, no.~6,
  pp.~1003--1016.

\end{thebibliography}


\begin{table*}[htbp]
  \centering
  \caption{Transition Model Features}
  \label{tab:transition_model_features}
    \begin{adjustbox}{width=\textwidth}
    \begin{tabular}{l|c|ccccc|ccccc}
    \toprule
    \multicolumn{1}{r}{} & \multicolumn{1}{c}{} & \multicolumn{5}{c}{\textbf{CAS Group}} & \multicolumn{5}{c}{\textbf{CAS$\rightarrow$RECOV}} \\
    \midrule
    \multicolumn{1}{l}{\textbf{Post Content}} & \multicolumn{1}{c}{p-value} & Mean  & Std.  & 25\%  & 50\%  & \multicolumn{1}{c}{75\%} & Mean  & Std.  & 25\%  & 50\%  & 75\% \\
    \midrule
    \multicolumn{1}{p{16.665em}|}{\% High Addiction Risk} & ***   & 0.807 & 0.184 & 0.706 & 0.846 & 0.944 & 0.113 & 0.191 & 0.000 & 0.000 & 0.167 \\
    \multicolumn{1}{p{16.665em}|}{\% Medium Addiction Risk} & *     & 0.103 & 0.138 & 0.000 & 0.063 & 0.143 & 0.113 & 0.191 & 0.000 & 0.000 & 0.167 \\
    \multicolumn{1}{p{16.665em}|}{\% Low Addiction Risk} & ***   & 0.091 & 0.136 & 0.000 & 0.045 & 0.129 & 0.038 & 0.126 & 0.000 & 0.000 & 0.000 \\
    Average Post Length & **    & 931.824 & 769.360 & 478.306 & 709.600 & 1145.870 & 816.774 & 630.207 & 446.318 & 653.833 & 977.833 \\
    \# Posts In First 6 Months \^ & ***   & 9.393 & 6.942 & 6.000 & 7.000 & 10.000 & 4.503 & 5.578 & 1.000 & 2.000 & 6.000 \\
    CAS $\rightarrow$ RECOV Keywords / \# Posts \^ & **    & 0.214 & 0.334 & 0.000 & 0.082 & 0.318 & 0.591 & 1.160 & 0.000 & 0.200 & 0.778 \\
    CAS Keywords / \# Posts \^ & ***   & 1.474 & 1.632 & 0.333 & 1.000 & 1.800 & 0.720 & 1.269 & 0.000 & 0.263 & 1.000 \\
    \midrule
    \multicolumn{1}{l}{\textbf{Drug Utterances}} & \multicolumn{1}{c}{p-value} & Mean  & Std.  & 25\%  & 50\%  & \multicolumn{1}{c}{75\%} & Mean  & Std.  & 25\%  & 50\%  & 75\% \\
    \midrule
    Buprenorphine \^ & ***   & 0.145 & 0.209 & 0.034 & 0.065 & 0.125 & 0.199 & 0.140 & 0.095 & 0.167 & 0.289 \\
    LSD \^ & ***   & 0.175 & 0.123 & 0.069 & 0.143 & 0.261 & 0.113 & 0.129 & 0.046 & 0.075 & 0.111 \\
    oxycodone \^ & ***   & 0.142 & 0.138 & 0.051 & 0.104 & 0.191 & 0.215 & 0.183 & 0.089 & 0.143 & 0.291 \\
    Heroin \^ & *     & 0.216 & 0.172 & 0.066 & 0.204 & 0.333 & 0.273 & 0.172 & 0.143 & 0.273 & 0.368 \\
    \bottomrule
    \end{tabular}\end{adjustbox}
\end{table*}

\begin{table*}[htbp]
  \centering
  \caption{Transition Model LIWC Features}
  \label{tab:transition_model_liwc}
    \begin{tabular}{l|c|ccccc|ccccc}
    \toprule
    \multicolumn{1}{r}{} & \multicolumn{1}{c}{} & \multicolumn{5}{c}{\textbf{CAS Group}} & \multicolumn{5}{c}{\textbf{CAS$\rightarrow$RECOV}} \\
    \midrule
    \multicolumn{1}{r}{} & \multicolumn{1}{c}{p-value} & Mean  & Std.  & 25\%  & 50\%  & \multicolumn{1}{c}{75\%} & Mean  & Std.  & 25\%  & 50\%  & 75\% \\
    \midrule
    leisure &  ***  & 1.109 & 0.992 & 0.516 & 0.918 & 1.502 & 0.706 & 0.759 & 0.263 & 0.541 & 0.988 \\
    Parenth &  ***  & 1.137 & 1.088 & 0.358 & 0.864 & 1.574 & 0.713 & 0.771 & 0.147 & 0.447 & 1.138 \\
    Sixltr &  ***  & 14.517 & 2.996 & 12.552 & 14.372 & 16.228 & 13.245 & 2.920 & 11.167 & 12.983 & 15.029 \\
    anx   &  ***  & 0.368 & 0.313 & 0.133 & 0.302 & 0.516 & 0.258 & 0.261 & 0.026 & 0.194 & 0.383 \\
    assent &  **   & 0.236 & 0.319 & 0.050 & 0.173 & 0.295 & 0.206 & 0.427 & 0.000 & 0.097 & 0.267 \\
    Comma &  **   & 3.522 & 1.945 & 2.172 & 3.373 & 4.586 & 2.826 & 1.686 & 1.478 & 2.844 & 3.993 \\
    insight &  **   & 3.017 & 1.147 & 2.289 & 2.860 & 3.582 & 2.609 & 0.999 & 1.913 & 2.546 & 3.246 \\
    informal &  **   & 1.823 & 1.526 & 0.916 & 1.442 & 2.220 & 2.099 & 1.325 & 1.133 & 1.873 & 2.804 \\
    Period &  **   & 5.593 & 2.284 & 4.224 & 5.449 & 6.543 & 6.146 & 2.110 & 4.848 & 6.043 & 7.031 \\
    netspeak &  **   & 0.667 & 0.756 & 0.172 & 0.461 & 0.934 & 0.879 & 0.812 & 0.277 & 0.700 & 1.266 \\
    swear &  **   & 0.551 & 0.676 & 0.148 & 0.361 & 0.678 & 0.656 & 0.594 & 0.202 & 0.565 & 0.883 \\
    anger &  *    & 0.688 & 0.856 & 0.233 & 0.513 & 0.810 & 0.780 & 0.607 & 0.296 & 0.695 & 1.139 \\
    focuspresent &  *    & 11.483 & 2.254 & 9.861 & 11.475 & 12.802 & 12.050 & 2.333 & 10.623 & 12.140 & 13.377 \\
    ingest &  *    & 0.954 & 0.762 & 0.493 & 0.808 & 1.211 & 0.782 & 0.650 & 0.304 & 0.678 & 1.064 \\
    time  &  *    & 5.857 & 1.591 & 4.732 & 5.839 & 6.933 & 6.330 & 1.998 & 4.804 & 6.288 & 7.794 \\
    cogproc &  *    & 14.827 & 2.808 & 13.224 & 14.942 & 16.450 & 14.265 & 2.951 & 12.346 & 14.057 & 15.963 \\
    reward &  *    & 2.004 & 0.944 & 1.433 & 1.876 & 2.406 & 2.188 & 0.912 & 1.542 & 2.108 & 2.569 \\
    motion &  *    & 1.672 & 0.673 & 1.266 & 1.688 & 2.068 & 1.549 & 0.772 & 1.078 & 1.484 & 1.999 \\
    prep  &  *    & 12.470 & 1.687 & 11.490 & 12.373 & 13.476 & 12.853 & 1.798 & 11.832 & 12.705 & 14.087 \\
    sad   & 0.053 & 0.337 & 0.311 & 0.116 & 0.264 & 0.469 & 0.424 & 0.462 & 0.184 & 0.338 & 0.535 \\
    money & 0.060 & 0.427 & 0.458 & 0.107 & 0.332 & 0.628 & 0.529 & 0.502 & 0.157 & 0.408 & 0.788 \\
    focuspast & 0.073 & 4.442 & 1.574 & 3.368 & 4.529 & 5.422 & 4.199 & 1.510 & 3.051 & 4.078 & 5.149 \\
    bio   & 0.073 & 3.867 & 1.498 & 2.794 & 3.819 & 4.735 & 3.586 & 1.345 & 2.718 & 3.470 & 4.338 \\
    WPS   & 0.079 & 16.701 & 4.792 & 13.820 & 15.546 & 18.528 & 16.381 & 6.760 & 12.846 & 15.092 & 18.087 \\
    health & 0.089 & 1.888 & 0.946 & 1.187 & 1.848 & 2.539 & 1.761 & 1.056 & 1.045 & 1.522 & 2.232 \\
    they  & 0.090 & 0.501 & 0.504 & 0.148 & 0.393 & 0.696 & 0.605 & 0.584 & 0.203 & 0.470 & 0.904 \\
    \bottomrule
    \end{tabular}%

\end{table*}
\begin{table*}[htbp]
  \centering
  \caption{Survival Model Drug Features}
    \begin{tabular}{l|c|ccccc|ccccc}
    \toprule
    \multicolumn{1}{r}{} & \multicolumn{1}{c}{} & \multicolumn{5}{c}{\textbf{Survived 12 Months}} & \multicolumn{5}{c}{\textbf{Did Not Survive 12 Months}} \\
    \midrule
    \multicolumn{1}{r}{} & \multicolumn{1}{c}{p-value} & Mean  & Std.  & 25\%  & 50\%  & \multicolumn{1}{c}{75\%} & Mean  & Std.  & 25\%  & 50\%  & 75\% \\
    \midrule
    Buprenorphine &  ***  & 0.0893 & 0.1275 & 0.0000 & 0.0204 & 0.1429 & 0.2289 & 0.1479 & 0.1017 & 0.2223 & 0.3162 \\
    LSD   &  ***  & 0.0404 & 0.0843 & 0.0000 & 0.0000 & 0.0400 & 0.1188 & 0.1176 & 0.0116 & 0.0991 & 0.1884 \\
    Heroin &  ***  & 0.0950 & 0.1168 & 0.0000 & 0.0513 & 0.1550 & 0.0233 & 0.0424 & 0.0000 & 0.0000 & 0.0334 \\
    Ecstasy &  ***  & 0.0691 & 0.1090 & 0.0000 & 0.0257 & 0.0909 & 0.0198 & 0.0322 & 0.0000 & 0.0044 & 0.0306 \\
    amphetamine &  ***  & 0.0276 & 0.0601 & 0.0000 & 0.0000 & 0.0283 & 0.0076 & 0.0355 & 0.0000 & 0.0000 & 0.0000 \\
    ketamine &  ***  & 0.0133 & 0.0560 & 0.0000 & 0.0000 & 0.0000 & 0.0251 & 0.0554 & 0.0000 & 0.0000 & 0.0208 \\
    inhalants &  ***  & 0.0298 & 0.0707 & 0.0000 & 0.0000 & 0.0265 & 0.0084 & 0.0293 & 0.0000 & 0.0000 & 0.0000 \\
    oxycodone &  ***  & 0.0894 & 0.1103 & 0.0213 & 0.0565 & 0.1107 & 0.1112 & 0.1060 & 0.0348 & 0.0834 & 0.1456 \\
    psilocybin &  **   & 0.0048 & 0.0192 & 0.0000 & 0.0000 & 0.0000 & 0.0027 & 0.0103 & 0.0000 & 0.0000 & 0.0000 \\
    over\_counter &  **   & 0.0312 & 0.0748 & 0.0000 & 0.0000 & 0.0228 & 0.0076 & 0.0246 & 0.0000 & 0.0000 & 0.0000 \\
    kratom &  **   & 0.0226 & 0.0530 & 0.0000 & 0.0000 & 0.0236 & 0.0070 & 0.0169 & 0.0000 & 0.0000 & 0.0000 \\
    methylphenidate &  *    & 0.0130 & 0.0404 & 0.0000 & 0.0000 & 0.0000 & 0.0332 & 0.0705 & 0.0000 & 0.0000 & 0.0321 \\
    alcohol &  *    & 0.0292 & 0.0490 & 0.0000 & 0.0085 & 0.0408 & 0.0190 & 0.0407 & 0.0000 & 0.0000 & 0.0210 \\
    benzodiazepines &  *    & 0.0005 & 0.0047 & 0.0000 & 0.0000 & 0.0000 & 0.0030 & 0.0163 & 0.0000 & 0.0000 & 0.0000 \\
    \end{tabular}%
  \label{tab:Survival Model LIWC}%
\end{table*}

\begin{table*}[htbp]
  \centering
  \caption{Survival Model LIWC Features}
  \begin{tabular}{l|c|ccccc|ccccc}
  \toprule
  \multicolumn{1}{r}{} & \multicolumn{1}{c}{} & \multicolumn{5}{c}{\textbf{Survived 12 Months}} & \multicolumn{5}{c}{\textbf{Did Not Survive 12 Months}} \\
  \midrule
  \multicolumn{1}{r}{} & \multicolumn{1}{c}{p-value} & Mean  & Std.  & 25\%  & 50\%  & \multicolumn{1}{c}{75\%} & Mean  & Std.  & 25\%  & 50\%  & 75\% \\
  \midrule
  leisure &  ***  & 1.070 & 0.600 & 0.646 & 0.977 & 1.371 & 0.714 & 0.390 & 0.433 & 0.636 & 0.909 \\
  ingest &  ***  & 1.021 & 0.590 & 0.613 & 0.923 & 1.340 & 0.745 & 0.395 & 0.485 & 0.653 & 0.931 \\
  Period &  ***  & 5.768 & 2.594 & 4.265 & 5.357 & 6.723 & 6.700 & 2.255 & 5.350 & 6.247 & 7.829 \\
  time  &  ***  & 5.894 & 1.365 & 5.012 & 5.851 & 6.738 & 6.578 & 1.384 & 5.542 & 6.522 & 7.468 \\
  focuspresent &  ***  & 11.859 & 1.716 & 10.771 & 11.849 & 12.856 & 12.526 & 1.561 & 11.488 & 12.562 & 13.534 \\
  relativ &  ***  & 13.510 & 2.122 & 12.195 & 13.484 & 14.792 & 14.290 & 2.003 & 13.127 & 14.246 & 15.592 \\
  informal &  ***  & 2.172 & 1.423 & 1.243 & 1.831 & 2.731 & 2.580 & 1.333 & 1.586 & 2.432 & 3.181 \\
  nonflu &  ***  & 0.162 & 0.171 & 0.059 & 0.126 & 0.222 & 0.217 & 0.181 & 0.099 & 0.175 & 0.281 \\
  netspeak &  ***  & 0.890 & 0.899 & 0.368 & 0.658 & 1.147 & 1.094 & 0.775 & 0.542 & 0.937 & 1.419 \\
  sad   &  ***  & 0.352 & 0.231 & 0.199 & 0.309 & 0.454 & 0.432 & 0.263 & 0.275 & 0.382 & 0.525 \\
  percept &  ***  & 2.768 & 0.757 & 2.270 & 2.692 & 3.207 & 2.524 & 0.626 & 2.102 & 2.447 & 2.883 \\
  WPS   &  ***  & 18.022 & 7.864 & 14.088 & 16.391 & 19.665 & 16.039 & 5.381 & 12.763 & 15.185 & 17.411 \\
  Authentic &  ***  & 67.714 & 11.134 & 61.327 & 68.676 & 75.226 & 71.432 & 9.774 & 65.455 & 72.193 & 78.747 \\
  money &  ***  & 0.476 & 0.366 & 0.225 & 0.395 & 0.633 & 0.591 & 0.481 & 0.298 & 0.522 & 0.784 \\
  Sixltr &  ***  & 13.681 & 2.664 & 11.839 & 13.391 & 15.162 & 12.889 & 2.178 & 11.303 & 12.684 & 14.056 \\
  Exclam &  ***  & 0.353 & 0.624 & 0.027 & 0.145 & 0.412 & 0.500 & 0.747 & 0.074 & 0.234 & 0.664 \\
  insight &  ***  & 2.800 & 0.821 & 2.269 & 2.782 & 3.288 & 2.597 & 0.712 & 2.092 & 2.500 & 2.971 \\
  death &  ***  & 0.116 & 0.130 & 0.025 & 0.080 & 0.160 & 0.142 & 0.133 & 0.049 & 0.115 & 0.189 \\
  conj  &  ***  & 7.569 & 1.191 & 6.799 & 7.573 & 8.327 & 7.257 & 1.028 & 6.522 & 7.250 & 8.000 \\
  family &  ***  & 0.151 & 0.190 & 0.032 & 0.095 & 0.206 & 0.187 & 0.187 & 0.054 & 0.133 & 0.247 \\
  power &  ***  & 1.900 & 0.583 & 1.548 & 1.838 & 2.199 & 2.009 & 0.457 & 1.719 & 1.953 & 2.241 \\
  anger &  ***  & 0.887 & 0.681 & 0.458 & 0.737 & 1.150 & 1.008 & 0.669 & 0.554 & 0.896 & 1.298 \\
  prep  &  **   & 12.265 & 1.334 & 11.466 & 12.270 & 13.173 & 12.620 & 1.194 & 11.834 & 12.485 & 13.428 \\
  article &  **   & 5.275 & 0.924 & 4.738 & 5.287 & 5.827 & 5.080 & 0.766 & 4.648 & 5.019 & 5.539 \\
  ppron &  **   & 10.161 & 1.834 & 9.024 & 10.199 & 11.279 & 10.521 & 1.670 & 9.431 & 10.600 & 11.658 \\
  negemo &  **   & 2.315 & 0.934 & 1.698 & 2.202 & 2.771 & 2.503 & 0.923 & 1.844 & 2.424 & 2.919 \\
  bio   &  **   & 4.023 & 1.131 & 3.301 & 3.933 & 4.684 & 3.765 & 0.980 & 3.119 & 3.687 & 4.372 \\
  AllPunc &  **   & 16.884 & 5.313 & 13.640 & 16.374 & 19.133 & 17.659 & 4.533 & 14.710 & 17.257 & 20.393 \\
  negate &  **   & 1.818 & 0.534 & 1.464 & 1.783 & 2.128 & 1.922 & 0.482 & 1.595 & 1.905 & 2.189 \\
  swear &  **   & 0.784 & 0.715 & 0.323 & 0.619 & 1.023 & 0.884 & 0.695 & 0.439 & 0.728 & 1.200 \\
  compare &  **   & 2.664 & 0.683 & 2.204 & 2.618 & 3.076 & 2.526 & 0.624 & 2.062 & 2.455 & 2.901 \\
  filler &  **   & 0.091 & 0.126 & 0.011 & 0.057 & 0.125 & 0.101 & 0.098 & 0.030 & 0.072 & 0.147 \\
  space &  **   & 5.982 & 1.062 & 5.304 & 5.921 & 6.633 & 6.175 & 1.007 & 5.496 & 6.150 & 6.782 \\
  anx   &  **   & 0.349 & 0.250 & 0.183 & 0.296 & 0.456 & 0.293 & 0.184 & 0.181 & 0.256 & 0.377 \\
  focusfuture &  **   & 1.322 & 0.492 & 0.993 & 1.265 & 1.590 & 1.398 & 0.429 & 1.072 & 1.381 & 1.617 \\
  drives &  *    & 6.433 & 1.200 & 5.748 & 6.358 & 6.989 & 6.598 & 0.978 & 5.969 & 6.496 & 7.181 \\
  adverb &  *    & 6.324 & 1.080 & 5.643 & 6.318 & 7.041 & 6.533 & 0.994 & 5.864 & 6.522 & 7.162 \\
  you   &  *    & 1.131 & 0.725 & 0.625 & 0.976 & 1.459 & 1.232 & 0.698 & 0.733 & 1.093 & 1.555 \\
  social &  *    & 6.102 & 2.134 & 4.692 & 5.915 & 7.318 & 6.381 & 1.876 & 5.033 & 6.321 & 7.509 \\
  see   &  *    & 0.753 & 0.405 & 0.483 & 0.698 & 0.928 & 0.686 & 0.356 & 0.459 & 0.612 & 0.890 \\
  affect &  *    & 5.531 & 1.354 & 4.706 & 5.374 & 6.206 & 5.823 & 1.519 & 4.829 & 5.575 & 6.564 \\
  pronoun &  *    & 16.197 & 2.067 & 14.931 & 16.277 & 17.484 & 16.463 & 1.940 & 15.128 & 16.726 & 17.842 \\
  Parenth &  *    & 0.893 & 0.832 & 0.348 & 0.706 & 1.203 & 0.747 & 0.621 & 0.305 & 0.603 & 0.984 \\
  certain &  *    & 1.570 & 0.533 & 1.244 & 1.513 & 1.836 & 1.629 & 0.475 & 1.298 & 1.637 & 1.879 \\
  female &  *    & 0.271 & 0.332 & 0.040 & 0.162 & 0.386 & 0.280 & 0.265 & 0.076 & 0.217 & 0.405 \\
  i     &  *    & 7.633 & 1.688 & 6.597 & 7.727 & 8.752 & 7.904 & 1.560 & 6.997 & 7.864 & 8.990 \\
  Dic   &  *    & 86.958 & 3.542 & 85.352 & 87.458 & 89.201 & 87.535 & 2.807 & 86.150 & 87.861 & 89.459 \\
  Dash  &  *    & 0.577 & 0.778 & 0.160 & 0.358 & 0.733 & 0.486 & 0.543 & 0.108 & 0.286 & 0.626 \\
  verb  &  *    & 18.501 & 1.919 & 17.362 & 18.562 & 19.714 & 18.797 & 1.721 & 17.645 & 18.824 & 19.990 \\
  \bottomrule
  \end{tabular}%

  \label{tab:addlabel}%
\end{table*}

\end{document}